# Análisis bibliométrico de la producción científica sobre el efecto COVID-19 en las Ciencias de la Información

Bibliometric Analysis of Scientific Production on the COVID-19 Effect in Information Sciences

Carmen Gálvez[1*] https://orcid.org/0000-0001-7454-1254

[1]Universidad de Granada. Granada, España.

[*]Autor para la correspondencia: cgalvez@ugr.es

**RESUMEN**

En este trabajo se analiza la producción científica sobre el efecto COVID-19 en el área de las Ciencias de la Información desde una perspectiva bibliométrica. Los objetivos se centraron en: 1) determinar los autores, países, instituciones y revistas más productivas; 2) identificar las fuentes que constituyen el núcleo de la producción científica; 3) examinar los manuscritos con mayor impacto y 4) visualizar la estructura temática y conceptual del dominio científico analizado. Para el análisis de los datos se utilizaron indicadores bibliométricos y técnicas de análisis factorial. Se recuperó un total de 1,175 publicaciones indexadas en la colección central de *Web of Science* (WoS) desde 2020 hasta 2022. Los resultados mostraron que los países más relevantes fueron Estados Unidos, Reino Unido, China y España. El núcleo de la producción científica estuvo formado por las publicaciones: *Journal of the American Medical Informatics Association*, *Profesional de la Información, Scientometrics* y *Journal of Health Communication*. Los documentos con mayor impacto se concentraron en los trabajos dedicados al análisis de la función de la telemedicina en la atención médica. La estructura conceptual mostró los principales frentes de investigación, tales como la función de la telesalud, las bibliotecas académicas y la alfabetización digital en la lucha contra la pandemia, el papel de las redes sociales en la crisis sanitaria, así como el problema de la desinformación y las noticias falsas.

## ABSTRACT

This paper analyzes the scientific production on the COVID-19 effect in the area of Information Sciences from a bibliometric perspective. The objectives focused on: 1) determining the most productive authors, countries, institutions and journals; 2) identifying the sources that constitute the core of scientific production; 3) examining the manuscripts with the greatest impact; and 4) visualizing the thematic and conceptual structure of the scientific domain analyzed. Bibliometric indicators and factor analysis techniques were used for data analysis. A total of 1,175 publications indexed in the Web of Science (WoS) core collection from 2020 to 2022 were retrieved. The results showed that the most relevant countries were the United States, United Kingdom, China and Spain. The core of the scientific production was formed by the publications: Journal of the American Medical Informatics Association, Information Professional, Scientometrics and Journal of Health Communication. The papers with the greatest impact were concentrated in those dedicated to the analysis of the role of telemedicine in medical care. The conceptual structure showed the main research fronts, such as the role of telehealth, academic libraries and digital literacy in the fight against the pandemic, the role of social networks in the health crisis, as well as the problem of misinformation and fake news.

# Introducción

La pandemia de la COVID-19 ha provocado una de las mayores crisis sanitarias mundiales de los últimos años.[1] La búsqueda de soluciones al problema de salud pública global ha impactado en muchos aspectos de nuestra sociedad; uno de ellos es la información y comunicación científica que en sus inicios se enfrentó a la falta de datos de estudios de respaldo. La comunidad científica ha reaccionado rápidamente a los desafíos médicos generados por la pandemia. Desde diversas disciplinas, los investigadores comenzaron a publicar sus hallazgos con el fin de compartir información para hacer frente a la crisis sanitaria, lo que provocó un crecimiento del conocimiento, sin precedentes.[2] El pico de las publicaciones posteriores al brote del coronavirus de 2019 fue inédito: en menos de cinco meses se indexaron más de 12 000 artículos de investigación y, en menos de siete, más de 30 000 artículos.[3] Los científicos de todo el mundo están realizando investigaciones a velocidades vertiginosas para abordar el problema.

A su vez, la necesidad de la reducción del tiempo de publicación para artículos sobre la COVID-19 ha acelerado el proceso de revisión por pares, tanto es así que se están llegando a cambiar las normas de publicación de los trabajos de investigación[4] (los servidores de *preprints*, donde los científicos publican manuscritos antes de la revisión por pares, se han visto desbordados de estudios que, posteriormente, se divulgan en las redes sociales y en plataformas, como *Twitter*). La enorme demanda para producir evidencias relacionadas con la COVID-19, que puedan proporcionar una base de conocimiento para la toma de decisiones clínicas efectivas o para conocer los principales frentes de investigación en este campo, se ha vuelto esencial. Además de la orientación de la investigación hacia temas epidemiológicos relacionados con la virología, la investigación sobre la COVID-19 parece haberse convertido en un punto focal de interés en muchas y diversas disciplinas académicas, como es el caso de las ciencias de la computación,[5,6] o como se aprecia en el caso de las investigaciones en bibliotecología y los estudios de la información.[7] Desde un enfoque bibliométrico, que es en el que se encuadra este estudio, son muchos las investigaciones dirigidas al crecimiento exponencial de la producción científica sobre esta enfermedad.[8,9,10,11,12,13,14,15,16]

En general, el análisis bibliométrico se dirige a los estudios que dan cuenta del ritmo creciente de la producción científica, de los frentes y tendencias de investigación, de las instituciones y países implicados o de los temas más relevantes.[17] Los estudios métricos se fundamentan, básicamente, en el análisis cuantitativo de los datos estadísticos extraídos de la literatura publicada, utilizando dos tipos de medidas básicas:[18] 1) indicadores unidimensionales, basados en técnicas estadísticas univariables, dedicadas a analizar o medir una única característica de los documentos seleccionados, sin tener en cuenta ningún vínculo que pudiera haber entre ellos; 2) indicadores multidimensionales, basados en técnicas estadísticas multivariantes, dedicadas a analizar o medir de forma simultánea diferentes unidades de análisis o variables observadas en los documentos seleccionados. Los análisis métricos están continuamente experimentando el desarrollo de nuevas herramientas de análisis y visualización de la información, que contribuyen al interés en realizar este tipo de investigaciones.

La finalidad de este trabajo fue aplicar las técnicas bibliométricas para analizar y visualizar el impacto de la producción científica sobre la enfermedad de la COVID-19 en el ámbito de conocimiento de la Biblioteconomía y las Ciencias de la Información (BCI) (*Library and Information Science* [LIS]). Este estudio pretende responder a las siguientes cuestiones:

- ¿Cuáles son los autores, países, instituciones y revistas más relevantes sobre la COVID-19 en el área BIC?

- ¿Qué fuentes forman el núcleo de la producción científica sobre la COVID-19 en el área BIC y cuáles son las fuentes más citadas y de mayor impacto?

- ¿Cuáles son los manuscritos o documentos más citados?

- ¿Qué temáticas constituyen la estructura conceptual de la producción científica de la COVID-19 dentro del área BIC?

## Métodos

La metodología se desarrolló en las siguientes etapas: 1) obtención de los datos; 2) identificación de los países e instituciones más relevantes; 3) identificación de las revistas que constituyen el núcleo del área estudiada y cuáles son las que tienen mayor impacto, aplicando indicadores unidimensionales; 4) identificación de los documentos más citados y

5) identificación de los principales frentes de investigación, aplicando indicadores multidimensionales.

La fuente de información fue la base de datos *Web of Science* (WoS) (*Clarivate Analytics, Philadelphia*, PA, USA). La elección de WoS se debió a que proporciona numerosas herramientas de análisis para procesar los datos y ofrece información de investigación altamente precisa. Dentro de la colección principal de WoS, los datos se obtuvieron de las bases de datos *Science Citation Index Expanded* (SCI-EXPANDED), *Social Sciences Citation Index* (SSCI), *Arts & Humanities Citation Index* (A&HCI) y *Emerging Sources Citation Index* (ESCI). La estrategia de búsqueda empleada consistió en seleccionar los campos: WC (*Web of Science Categories* o búsqueda por las categorías de materia establecidas por WoS), TS (*Topic* o búsqueda por el tema en los campos título, resumen, palabras clave, *KeyWords Plus* y palabras clave de autor, *Author's Keywords*). La ecuación de búsqueda fue la siguiente: *(WC= Information Science & Library Science) AND (TS=COVID-19 OR 2019-nCoV OR SARS-CoV-2 OR new coronavirus OR coronavirus disease 2019 OR severe acute respiratory syndrome coronavirus 2)*. La indagación en WoS se realizó el 31 de enero de 2022.

Para identificar las revistas más productivas y los artículos de mayor impacto se aplicaron indicadores unidimensionales, utilizando dos medidas: a) indicadores de dispersión para determinar las publicaciones que constituyen el núcleo del área estudiada y b) indicadores de impacto para medir la influencia de los trabajos publicados en el campo. Dentro de los primeros indicadores, se aplicó el modelo de Bradford,[19] conocido como la ley bibliométrica de dispersión de la literatura, con la finalidad de conocer las revistas más productivas, que se encuentran a su vez en el núcleo de las revistas más especializas.

La ley bibliométrica de dispersión de la literatura establece la siguiente premisa: si las revistas científicas se ordenaran en una secuencia decreciente de productividad de artículos sobre un tema dado, estas podrían dividirse en un núcleo de revistas dedicadas más específicamente a ese tema. Esta ley trata de demostrar que en la producción de artículos en las revistas seleccionadas existe una distribución altamente desigual, donde la mayoría de los artículos están concentrados en una pequeña población de revistas, mientras que una pequeña proporción de artículos se dispersa sobre una alta cantidad de revistas. Con la aplicación de este indicador se pretendió obtener la relación cuantitativa entre las revistas y los artículos científicos contenidos en los datos de la producción científica seleccionada.

Por otra parte, se aplicó otro indicador unidimensional de impacto: el índice H,[20] como una medida para evaluar la influencia de las revistas científicas más relevantes. Con la aplicación de esta métrica se ordenan las publicaciones de mayor a menor, según el número de citas que hayan recibido. Su cálculo es sencillo, consiste en medir el número n de publicaciones citadas al menos n veces en un período de referencia (el índice H estaría constituido por el número en el que coinciden el número de orden con el número de citas). Este índice toma en cuenta tanto la productividad (número de publicaciones) como la citación y se aplica de forma independiente tanto a las métricas de los autores, como a otras métricas dirigidas a las revistas o a la producción científica de los países. Para el cálculo de las revistas, estas se ordenan y enumeran de forma descendente por su número de publicaciones (NP) y por el número total de citas (TC) recibidas (una revista tiene un índice H, si ha publicado H artículos con al menos H citas cada uno).

Para identificar la estructura conceptual y temática del campo, se utilizaron indicadores bibliométricos multidimensionales a través de dos procedimientos: a) análisis factorial, como técnicas de reducción de la dimensionalidad de los datos y b) técnicas de agrupación (*clustering*) para identificar los grupos de documentos que expresen los conceptos comunes.

Como método de reducción de la dimensionalidad, se realizó un análisis factorial,[21] que consiste en un procedimiento estadístico de reducción de datos capaz de sintetizar una gran cantidad de información en un número reducido de dimensiones. Aunque este tipo de análisis está basado en métodos algebraicos y estadísticos muy complejos, se trata de una técnica exploratoria muy intuitiva. El análisis factorial permitió representar cada uno de los valores posibles, de cada una de los conceptos seleccionadas, en un plano donde la posición relativa de cada concepto representó el grado de asociación entre ellos. Como sucede con otras técnicas estadísticas multivariantes de reducción de dimensiones (como el escalamiento multidimensional, el análisis de correspondencia o el análisis de correspondencia múltiple) se consiguió hacer una fotografía de una realidad compleja y multidimensional, capaz de representarse en un plano o mapa bidimensional. En concreto, se trasladó una nube de puntos o conceptos, definida en un espacio de muchas dimensiones, a un espacio de dos dimensiones donde se pudo visualizar la posición relativa de cada uno de esos conceptos. Previamente, se seleccionaron los conceptos a partir de las palabras clave de autor (*Author's Keywords*)[22] y se limitó en 50 el número de conceptos a representar.

A continuación, se realizó un análisis para clasificar los conceptos en grupos/conglomerados (clúster) que fueran lo más homogéneos posibles dentro de cada grupo y, a su vez, heterogéneos entre sí. La estrategia para definir los grupos se basó en el análisis de clúster k-medias (*k-means*)[23] o análisis de conglomerados no jerárquicos. El análisis clúster de k-medias es un procedimiento diseñado para asignar los casos a un número fijo de grupos; su objetivo es maximizar la homogeneidad dentro de cada grupo. No obstante, este tipo de análisis requiere conocer el número exacto de clústeres, por eso la elección correcta del número de grupos (k) puede llegar a ser ambigua y compleja. En este trabajo fue necesario encontrar un equilibrio entre el conjunto de datos analizados y el número de clústeres que se deseaba obtener. Finalmente, el número de clústeres o agrupamientos se fijó en cinco (k = 5). Con la aplicación de los indicadores multidimensionales se obtuvo la estructura conceptual y temática del área de estudio, que proporcionó los principales frentes de investigación.

Para el procesamiento estadístico y descriptivo de los datos, así como para la identificación conceptual, se empleó el paquete R Bibliometrix versión 4.1.0 y la aplicación Biblioshiny (interfaz *web* de Bibliometrix).[24] Se trata de una herramienta de código abierto, desarrollada en R, que incluye los principales métodos de análisis bibliométricos. El paquete permite importar datos bibliográficos de las principales bases de datos científicas como Scopus, WoS y PubMed, entre otras.

## Resultados y discusión

### Información general sobre la colección

A partir de los criterios de búsqueda seleccionados se obtuvo un total de 1175 publicaciones en la base de datos WoS. Sus características específicas se presentan en la tabla 1: 828 artículos, 3409 palabras clave de autor (DE) y 3577 autores, de los cuales 268 documentos contaron con autores únicos y 3309 con autoría múltiple.

**Tabla 1** - Información global de los datos obtenidos de *Web of Science* (WoS)

| Descripción | Resultados |
|---|---|
| **Información principal sobre los datos** | **2020-2022** |
| Fuentes (revistas, libros, etc.) | 136 |
| Documentos | 1175 |
| **Tipos de documentos** | |
| Artículos | 828 |
| Reseña de libros | 6 |
| Cartas | 12 |
| Actas de congreso | 49 |
| Revisiones | 25 |
| **Contenidos de los documentos** | |
| Palabras clave (*Keywords Plus*, ID) | 1166 |
| Palabras clave de autor (*Author's Keywords*, DE) | 3408 |
| **Autores** | |
| Autores | 3577 |
| Apariciones de los autores | 4015 |
| Documentos de un solo autor | 268 |
| Documentos de varios autores | 3309 |

*Fuente*: Elaboración propia.

## Autores, países e instituciones más relevantes

Los autores más relevantes se obtuvieron a partir del número de artículos publicados por cada investigador (tabla 2). En las primeras posiciones se encontraron las contribuciones de los investigadores Thelwall (Universidad de Wolverhampto, Reino Unido), seguido de Xu (Universidad de Texas, USA) y Mahmood (Universidad del Panyab, Pakistán) todos con seis artículos publicados. También en las primeras posiciones se situaron las contribuciones de los investigadores Michalak (Goldey-Beacom College, USA), Rysavy (Goldey-Beacom College, USA), Scoulas (Universidad de Illinois, USA) con cinco artículos publicados.

**Tabla 2** - Clasificación de los 10 principales autores según el número de artículos publicados

| Rank | Autores | Afiliación | Artículos |
|---|---|---|---|
| 1 | Thelwall, Mike | Universidad de Wolverhampton, Reino Unido | 6 |
| 2 | Xu, Hua | Universidad de Texas, USA | 6 |
| 3 | Mahmood, Khalid | Universidad del Panyab, Pakistán | 6 |
| 4 | Michalak, Russell | Goldey–Beacom College, USA | 5 |
| 5 | Rysavy, Monica DT | Goldey–Beacom College, USA | 5 |

| 6 | Scoulas, Jung Mi | Universidad de Illinois, USA | 5 |
| 7 | Feng, Xin | Universidad de Yanshan, China | 4 |
| 8 | Jiang, Xiaoqian | Universidad de Texas, USA | 4 |
| 9 | Lenert, Leslie A. | Universidad Médica de Carolina del Sur, USA | 4 |
| 10 | Montesi, Michela | Universidad Complutense de Madrid, España | 4 |

*Fuente*: Elaboración propia.

En la figura 1 se muestra la visualización de los países más relevantes, según el origen de los autores de la correspondencia. El autor de la correspondencia se considera a la persona que actúa como el representante del resto de coautores y establece el contacto con el editor en jefe y los editores asociados de la revista en particular, durante el proceso de envío, revisión y edición final del manuscrito). El índice MCP (*Multiple-Country Publications*) representa el índice de colaboración interpaíses e indica el número de documentos en los que hay, al menos, un coautor de un país diferente. El índice SCP (*Single-Country Publications*) representa el índice de colaboración entre países e indica el número de documentos en los que todos los coautores del manuscrito tienen la afiliación del mismo país.

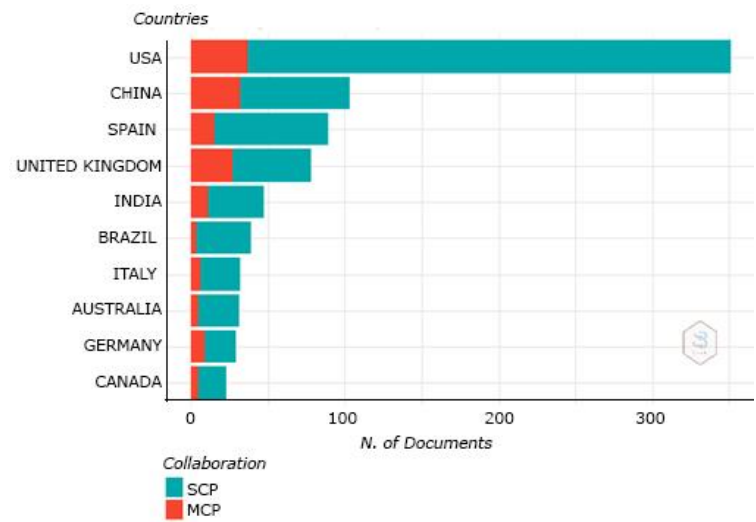

Leyenda: Índices de colaboración: SCP (*Single-Country Publications* o publicaciones de un solo país) y MCP (*Multiple-Country Publications* o publicaciones de múltiples países).

*Fuente:* Elaboración propia.

**Fig. 1** – Visualización de los 10 primeros países según el número de publicaciones del autor de la correspondencia.

En la tabla 3 se muestra la clasificación de los principales países, según el país de la correspondencia del primer autor: USA destaca del resto con un total de 351 autores de la correspondencia y una frecuencia de 30 participaciones; seguida de China, con un total de 103 autores y una frecuencia de nueve participaciones y, en tercer lugar, España con un total de 89 autores y una frecuencia de 7,6 participaciones.

**Tabla 3** - Clasificación de los 10 principales países, según el país de la correspondencia del primer autor

| Rank | País | Artículos | Frecuencia | Publicaciones de un solo país (SCP) | Publicaciones de múltiples países (MCP) |
|---|---|---|---|---|---|
| 1 | USA | 351 | 30 | 314 | 37 |
| 2 | China | 103 | 9 | 71 | 32 |
| 3 | España | 89 | 7,6 | 74 | 15 |
| 4 | Reino Unido | 78 | 6,7 | 51 | 27 |
| 5 | India | 47 | 4 | 36 | 11 |
| 6 | Brasil | 39 | 3,3 | 36 | 3 |
| 7 | Italia | 32 | 2,7 | 26 | 6 |
| 8 | Australia | 31 | 2,6 | 26 | 5 |
| 9 | Alemania | 29 | 2,5 | 20 | 9 |
| 10 | Canadá | 23 | 2 | 18 | 5 |

Leyenda: Índices de colaboración: SCP (*Single-Country Publications* o publicaciones de un solo país) y MCP (*Multiple-Country Publications* o publicaciones de múltiples países).

En la tabla 4 se muestran las afiliaciones más relevantes, según la distribución de frecuencia de afiliaciones de todos los coautores de cada artículo. Además, para comparar los resultados se utilizó uno de los *rankings* internacionales de universidades de mayor impacto: *Ranking de Shanghái* o *ARWU World University y Rankings* 2021,[25] dentro de la categoría académica LIS (*Library & Information Science*).

La Universidad de Columbia fue la que encabezó la clasificación con el mayor volumen de artículos (38), seguida de la Universidad Médica de Carolina del Sur (*Medical University of South Carolina*) con 23 artículos. En tercera posición se encuentra la Universidad de Colorado en Denver (23 artículos); en cuarta posición, la Universidad Vanderbil (20 artículos); la Universidad de Michigan en Ann Arbor (19 artículos), en quinta posición y la Universidad de California en San Francisco (18 artículos) en sexta posición. El resto de instituciones también son estadounidenses, a excepción de la Universidad de Wuhan, en séptima posición con 17 artículos. La mayoría de los centros presentes en las afiliaciones más importantes ocupan posiciones relevantes en la clasificación ARWU, con excepción de la Universidad Médica de Carolina del Sur y la Universidad de Colorado, debido a que están incluidas dentro de la categoría LIS en ARWU.

**Tabla 4** - Clasificación de las 10 afiliaciones más relevantes

| Rank | Organización | País | Artículos | ARWU 2021 en LIS |
|---|---|---|---|---|
| 1 | Universidad de Columbia | USA | 38 | 12 |
| 2 | Universidad Médica de Carolina del Sur | USA | 23 | - |
| 3 | Universidad de Colorado, Denver | USA | 23 | - |
| 4 | Universidad Vanderbil | USA | 20 | 4 |
| 5 | Universidad de Míchigan, Ann Arbor | USA | 19 | 5 |
| 6 | Universidad de California, San Francisco | USA | 18 | 51-75 |
| 7 | Universidad de Wuhan | China | 17 | 7 |
| 8 | Universidad de Indiana, Bloomington | USA | 16 | 3 |
| 9 | Universidad Stanford | USA | 16 | 21 |
| 10 | Universidad de Illinois, Urbana-Champaign | USA | 15 | 76-100 |

Nota al pie: La clasificación se realiza, según la según la distribución de frecuencia de afiliaciones de todos los coautores de cada artículo y su respectiva posición en *Ranking de Shanghái* o ARWU 2021 (*Academic Ranking of World Universities*, 2021) dentro de la categoría LIS (*Library & Information Science*).

*Fuente*: Elaboración propia.

## Revistas más relevantes y de mayor impacto

Atendiendo a la ley de Bradford,[19] las revistas que constituyen el núcleo del área estudiada se concentraron en nueve publicaciones en orden decreciente de productividad (fig. 2): *Journal of the American Medical Informatics Association* (113 artículos), *Profesional de la Información* (60 artículos), *Scientometrics* (43 artículos), *Journal of Health Communication* (36 artículos), *International Journal of Information Management* (34 artículos)*, Library Hi Tech* (31 artículos), *Information and Learning Sciences* (28 artículos)*, Information Processing & Management* (28 artículos) y *European Journal of Information Systems* (27 artículos).

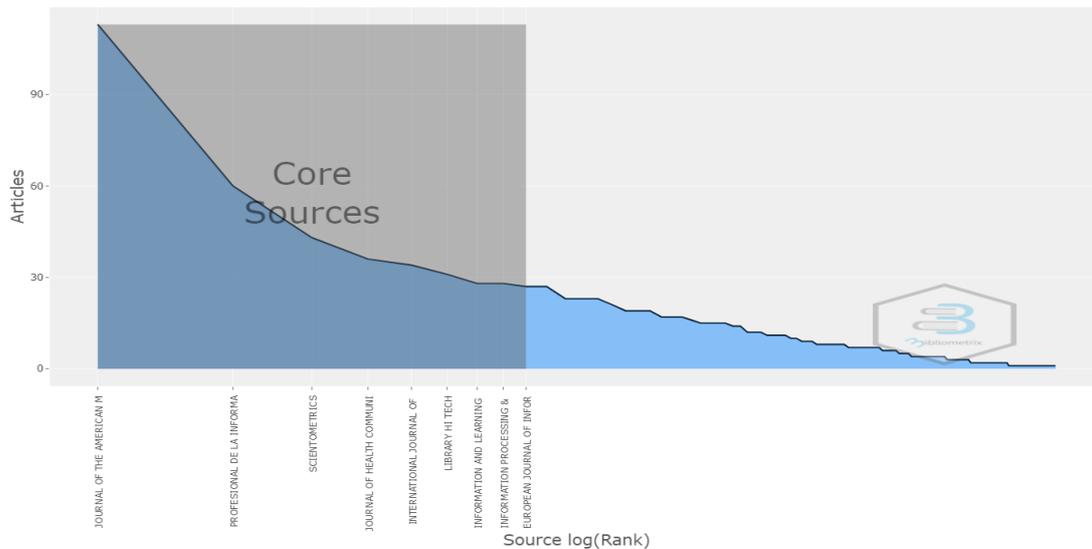

*Fuente:* Elaboración propia.

**Fig. 2** – Núcleo de revistas, en orden decreciente de productividad de artículos.

La tabla 5 muestra las fuentes más relevantes e influyentes según el índice, el número total de citas (TC), el número de publicaciones (NP) y el año inicial de la publicación de los manuscritos sobre el tema COVID-19 en las fuentes. La fuente más relevante fue, en primer lugar, la *International Journal of Information Management*, con un índice H de 21 y 1203 citas en total, seguida por la *Journal of the American Medical Informatics Association,* con un índice H de 18 y un total de 1879 citas; en tercera posición se encuentra el *Profesional de la Información* con 691 citas; en cuarto lugar, la revista *European Journal of Information Systems* (504 citas) y en quinto lugar la *Scientometrics* con 308 citas.

**Tabla 5 -** Clasificación de las 10 publicaciones más relevantes y su impacto según el índice H

| Rank | Fuente | Índice-H | TC | NP | Año |
|---|---|---|---|---|---|
| 1 | *International Journal of Information Management* | 21 | 1203 | 31 | 2020 |
| 2 | *Journal of the American Medical Informatics Association* | 18 | 1879 | 88 | 2020 |
| 3 | *Profesional de la Información* | 15 | 691 | 40 | 2020 |
| 4 | *European Journal of Information Systems* | 12 | 504 | 26 | 2020 |
| 5 | *Scientometrics* | 9 | 308 | 33 | 2020 |
| 6 | *Information and Learning Sciences* | 8 | 194 | 26 | 2020 |
| 7 | *Journal of Health Communication* | 6 | 125 | 21 | 2020 |

| 8 | *Ethics and Information Technology* | 6 | 114 | 12 | 2021 |
| 9 | *Government Information Quarterly* | 5 | 99 | 7 | 2020 |
| 10 | *Information Processing & Management* | 5 | 92 | 14 | 2021 |

Leyenda: TC = número total de citas (*total citations*); NP = número de publicaciones (*number of publications*).

Nota al pie: el año se corresponde al inicio de la publicación de los manuscritos sobre el tema COVID-19 en las fuentes.

*Fuente*: Elaboración propia.

## Documentos más relevantes

La tabla 6 muestra la lista de los manuscritos más relevantes ordenados por el número de citas recibidas en la muestra seleccionada, a partir de los documentos indexados en una base de datos WoS. El manuscrito más citado fue *Telehealth transformation: COVID-19 and the rise of virtual care*[26] que es un estudio sobre el uso de las tecnologías de la comunicación en la atención médica a distancia y muestra la importante transformación que ha experimentado la telesalud y la telemedicina durante la pandemia. Cuenta con 392 citas y una citación total por año de 130 667. Otro artículo muy citado fue *COVID-19 transforms health care through telemedicine: Evidence from the field*,[27] una investigación que aborda el impulso y el cambio que ha experimentado la telemedicina en la atención médica; obtuvo 335 citas y una citación total por año de 111 667.

**Tabla 6 -** Clasificación de los 10 artículos más citados

| Rank | Artículo | DOI | Referencia | Citas Totales (TC) | TC por Año |
| --- | --- | --- | --- | --- | --- |
| 1 | *Wosik J, 2020, J Am Med Inform Assn* | 10.1093/jamia/ocaa067 | [26] | 392 | 130,667 |
| 2 | *Mann DM, 2020, J Am Med Inform Assn* | 10.1093/jamia/ocaa072 | [27] | 335 | 111,667 |
| 3 | *Reeves JJ, 2020, J Am Med Inform Assn* | 10.1093/jamia/ocaa037 | [28] | 153 | 51 |
| 4 | *Dwivedi YK, 2020, Int J Inform Manage* | 10.1016/j.ijinfomgt.2020.102211 | [29] | 139 | 46,333 |
| 5 | *Casero-Ripolles A, 2020, Prof Inform* | 10.3145/epi.2020.mar.23 | [30] | 133 | 44,333 |

| 6 | Livari N, 2020, Int J Inform Manage | 10.1016/j.ijinfomgt.2020.102183 | [31] | 114 | 38 |
| 7 | Ramsetty A, 2020, J Am Med Inform Assn | 10.1093/jamia/ocaa078 | [32] | 108 | 36 |
| 8 | Laato S, 2020, Eur J Inform Syst | 10.1080/0960085X.2020.1770632 | [33] | 98 | 32,667 |
| 9 | De R, 2020, Int J Inform Manage | 10.1016/j.ijinfomgt.2020.102171 | [34] | 93 | 31 |
| 10 | Apuke OD, 2021, Telemat Inform | 10.1016/j.tele.2020.101475 | [35] | 86 | 43 |

Leyenda: TC = número total de citas (*total citations*); TC = número total de citas por año.

*Fuente*: Elaboración propia.

## Estructura conceptual y temática

El resultado la estructura conceptual y temática, obtenido a partir de la aplicación del análisis factorial y técnicas de reducción de datos, así como análisis de clúster k-medias, en el que k = 5, fue un mapa bidimensional en el que se reflejaron las asociaciones homogéneas, tanto de las palabras clave de autor (*author's keywords*) como de los artículos seleccionados. En la figura 3 se muestran las cinco agrupaciones obtenidas.

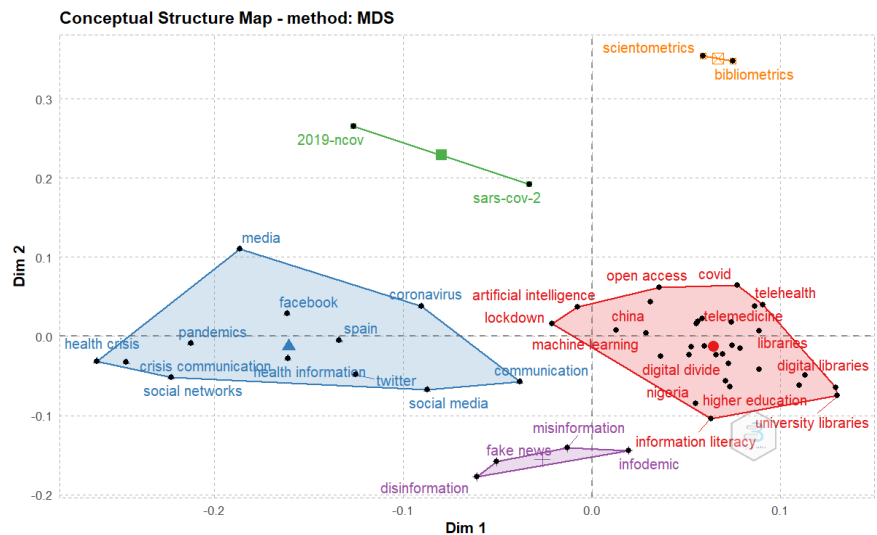

*Fuente:* Elaboración propia.

**Fig. 3** – Mapa de la estructura conceptual del campo analizado y su visualización, aplicando técnicas de reducción de datos MDS (*Multidimensional Scaling*).

Los cinco ejes temáticos que configuraron la estructura conceptual se describen a continuación (dos de ellos se mostraron bien cohesionados y otros tres reflejaron baja densidad):

- Clúster rojo: Muestra una alta concentración de conceptos, aunque los términos que lo integran son heterogéneos. Este clúster agrupó palabras clave relacionadas con la transformación en la que diversos sectores se enfrentaron a la gestión de información durante la pandemia y que determinaron un cambio sin precedentes en la transición de los servicios de modo físico a modo en línea. La crisis del coronavirus ha generado diferentes iniciativas para promover el acceso abierto (*open access*) a las publicaciones y a los datos, con la finalidad de resolver de manera colaborativa el problema sanitario, tales como el fomento de la ciencia abierta o el aprendizaje en línea (*machine learnnig*). La pandemia también ha catalizado la rápida adopción de la telesalud (*telehealth*) y ha impulsado una rápida expansión del uso de la telemedicina (*telemedicine*) en la prestación de atención médica. En este grupo también destacaron términos como bibliotecas digitales (*digital libraries*), bibliotecas académicas (*university libraries*), brecha digital (*digital divide*), habilidades de alfabetización digital (*information literacy*), implicación de la educación superior (*higher education*) en la comunicación, enseñanza y competencias digitales. La palabra clave *Nigeria* aparece próxima a educación superior por los estudios que vinculan el papel de las bibliotecas en los países subdesarrollados o en vías de desarrollo (como India y Pakistán) en la difusión de información sobre la COVID-19.

- Clúster azul: Presenta una alta concentración de conceptos, con homogeneidad semántica. Incluyó como término central información médica (*health information*), asociado a palabras como redes sociales (*social networks*), medios de comunicación social (*social media*), crisis comunicación (*crisis communication*), *Facebook* o *Twitter*. El bloque conceptual y temático de este grupo se centró en el papel fundamental de las redes sociales como canales directos de difusión de la información médica. Este clúster reflejó cómo la pandemia de la COVID-19 fue una crisis sanitaria que se extendió a escala global, también a través de las redes sociales. Plataformas como *Facebook*, *Twitter* se convirtieron en altavoces de todo tipo de información sobre el coronavirus y son muchos las investigaciones que trataron este tema.

- Clúster púrpura: En este grupo destacan palabras como noticias falsas (*fake news*), desinformación (*disinformation*), información errónea (*misinformation*) y sobreabundancia de información o infomedia (*infomedic*). Si bien las redes sociales se han utilizado para la difusión de la información médica, también se han empleado para distribuir noticias falsas y rumores que nadie ha confirmado ni verificado. Este grupo temático muestra, además, el problema que constituye la sobreabundancia de información o infodemia, que es capaz de engañar a las personas, hacer perder la confianza en el gobierno y las autoridades reguladoras de la salud.

- Clúster verde (dos palabras clave de autor): Esta agrupación muestra el gran impulso de la producción académica sobre el nuevo coronavirus, a través de la conexión entre las palabras clave que identifican la aparición de la nueva enfermedad (SARS-CoV-2) y el agente causal que la provocó (2019-ncov). El nuevo coronavirus pertenece a la especie del síndrome respiratorio agudo severo coronavirus 2 (SARS-CoV-2). En el año 2019, cuando la OMS recibió información sobre casos de neumonía en la ciudad china de Wuhan, no se pudo conocer la causa de la infección. Sin embargo, en el 2020 los investigadores médicos de China revelaron el misterio en torno al agente causal que provocó la infección (2019-ncov). Este hallazgo condujo a una proliferación de estudios en muchos ámbitos del conocimiento, lo que evidenció, probablemente, el caso más excepcional de estallido de publicaciones en la historia de la ciencia.

- Clúster amarillo: Este clúster muestra las dimensiones cienciométricas de la investigación del nuevo coronavirus, utilizando características cuantificables del conjunto de datos de las publicaciones científicas. Este clúster señala la conexión de las palabras clave bibliometría (*bibliometrics*) y cienciometría (*scientometrics*) para ilustrar este fenómeno excepcional en las publicaciones académicas y estudiar el enorme impacto de la explosión de publicaciones que utilizan los indicadores bibliométricos.

# Conclusiones

En un siglo globalizado e interconectado como el que nos ocupa los efectos de la crisis sanitaria han estado estrechamente vinculados a la economía, la política y, sin lugar a dudas, a la gestión de la información médica. Cuando la Organización Mundial de la Salud (OMS) declaró la pandemia de la COVID-19 se generaron cambios repentinos en la forma de transmitir la información y las habilidades informacionales que determinan la capacidad de los individuos para obtener información creíble y útil para ellos.

La pandemia ha dado lugar a la generación de una gran cantidad de publicaciones científicas, lo que ha propiciado problemas con la velocidad y la disponibilidad de la información. El objeto de este estudio ha sido el análisis bibliométrico de la producción científica sobre el efecto de la pandemia en el área de la Biblioteconomía, las Ciencias de la Documentación e Información, en función de poner de manifiesto cómo se ha enfrentado esta área de conocimiento a los retos planteados por la enfermedad.

Para ello, se ha intentado destacar a los investigadores, países e instituciones con mayor número de publicaciones y los trabajos con mayor impacto, así como los frentes de investigación más relevantes y con mayor número de publicaciones sobre el tema. Los investigadores que destacaron son: *Thelwall* (Universidad de Wolverhampton, Reino Unido), *Xu* (Universidad de Texas, USA) y *Mahmood* (Universidad del Panyab, Pakistán). En el caso de los países con mayor número de trabajos dedicados a este tema, sobresale Estados Unidos (USA), seguido de China y España. También las instituciones con mayor número de contribuciones pertenecieron a USA, tales como la Universidad de Columbia, seguida de la Universidad Médica de Carolina del Sur y la Universidad de Colorado en Denver.

Por otra parte, los manuscritos más relevantes han estado relacionados con el uso de las tecnologías de la comunicación y la transformación experimentada por la telemedicina en la atención primaria. Con respecto a la estructura conceptual, el análisis de co-palabras mostró los principales frentes de investigación, tales como los estudios sobre la función de las nuevas tecnologías de la comunicación, la telesalud y telemedicina en la atención primaria; las funciones de las bibliotecas académicas y la alfabetización digital en la lucha contra la pandemia; los estudios sobre la función de las redes sociales y los nuevos medios de comunicación para hacer frente a la crisis sanitaria provocada por el coronavirus; las investigaciones relacionadas con los estudios sobre el problema de la desinformación; la infomedia y las noticias falsas, así como los trabajos con un enfoque bibliométrico para analizar la explosión de publicaciones provocadas por la pandemia.

No obstante, es importante destacar que la rápida evolución de la pandemia y el incremento exponencial de la producción científica, junto con los nuevos resultados de investigación, representarían una limitación al presente estudio. Otra limitación sería que solo se ha analizado la información extraída de la base de datos WoS en idioma inglés, lo que podría causar cierto sesgo en los resultados. Por lo tanto, estudios futuros deberían abordar estas limitaciones y mostrar el progreso en el área de conocimiento analizada.

# Referencias bibliográficas

**Conflicto de intereses**

La autora declara que no tiene conflicto de intereses.